\documentclass[fleqn,10pt]{wlscirep}

\newcommand{\fig}[1]{Fig.~\ref{#1}}
\newcommand{\be}[1]{\begin{equation}\label{#1}}
\newcommand{\ee}{\end{equation}}
\usepackage{amsmath}
\usepackage{array}
\usepackage{setspace} 
\usepackage{float} 
\usepackage{tabularx}

\title{Fingerprints of slingshot non-sequential double ionization on two-electron probability distributions}

\author[1]{G. P. Katsoulis}
\author[1]{A. Emmanouilidou}
\affil[1]{Department of Physics and Astronomy, University College London, Gower Street, London WC1E 6BT, United Kingdom}

\affil[*]{a.emmanouilidou@ucl.ac.uk}


\begin{abstract}
We study double ionization of He driven by a near-single-cycle laser pulse at low intensities at 400 nm. Using a three-dimensional semiclassical model, we identify the pathways that prevail non-sequential double ionization (NSDI). We  focus mostly on the delayed pathway, where one electron ionizes with a time-delay after recollision.
We have recently shown that the mechanism that prevails the delayed pathway depends on intensity. For low intensities  slingshot-NSDI is the mechanism that prevails. Here, we identify the differences in  two-electron probability distributions of the prevailing NSDI pathways. This allows us to identify properties of the two-electron escape and thus gain significant insight into slingshot-NSDI.  Interestingly,  we find that an observable  fingerprint of slingshot-NSDI is the two electrons escaping with large and roughly equal energies.
 \end{abstract}
\begin{document}

\flushbottom
\maketitle
%
%
\thispagestyle{empty}

\section*{Introduction}
Non-sequential double ionization (NSDI) of atoms driven by intense laser fields is a fundamental process which has attracted considerable theoretical and experimental interest \cite{Corkum,PhysRevA.27.2503,PhysRevLett.55.2141,PhysRevLett.73.1227,0953-4075-31-6-008,PhysRevLett.84.447,RESI1,Weber,RESI2,DORNER20021,kinematic1,PhysRevLett.92.173001,PhysRevLett.93.253001,PhysRevLett94.093002,NSDI2,Taylor1,vshape2,vshape1,NSDI1,PfeifferNature,Dorner2013,Chen2016,Dorner2018}. The three-step model underlies NSDI  \cite{Corkum}. First, one electron tunnels through the field-lowered Coulomb potential. This tunnel-ionizing electron can return to the core and transfer energy to the other electron. For high intensities, the main pathway of NSDI  is the direct one.  The recolliding electron transfers  sufficient  energy to the other electron  for both electrons to ionize shortly after recollision. For low intensities, the delayed pathway of NSDI prevails. The recolliding electron transfers enough energy to the bound electron  for only one of the two electrons to ionize shortly after recollision. The other electron transitions to an exited state of the remaining ion. 

Until recently, the delayed pathway of NSDI was generally accepted as being equivalent to recollision-induced excitation with subsequent field ionization (RESI)  \cite{RESI1,RESI2}. According to RESI, the electron that transitions to an excited state after recollision, ionizes later in time at extrema of the laser field. It does so  mainly with the assistance of the field.  However, recently, we have shown that RESI does not always prevail the delayed pathway of NSDI. For He driven at 400 nm by near-single-cycle pulses, at intensities below the recollision threshold, we have shown  that a new mechanism overtakes RESI in the delayed pathway  of NSDI \cite{AgapiSlingshot}. We labeled this mechanism  slingshot-NSDI. We have shown that in slingshot-NSDI, following the transition to an excited state, 
 the electron  subsequently  undergoes a Coulomb slingshot motion due to the attractive force of the nucleus. Coulomb slingshot is similar to the
gravitational slingshot motion that alters the motion of a spacecraft around a planet. During Coulomb slingshot the electron ionizes mostly around the second extremum of the field with the assistance of both the nucleus and the laser field. Due to the  electron that ionizes last undergoing  slingshot motion, the two electrons escape in opposite directions along the laser field. 
Owning to this anti-correlated two-electron escape, slingshot-NSDI offers an alternative explanation to multiple recollisions.  Multiple recollisions is a mechanism that was previously put forth in the context of RESI  to explain anti-correlated two-electron escape  \cite{Anticor1, Anticor2, Anticor4,Anticor8, Anticor7}. This two-electron escape pattern   has been  found to prevail NSDI of several atoms driven by intense long duration pulses  and has been the object of many theoretical and experimental studies \cite{Anticor1,Anticor2,Anticor4,Anticor8,Anticor7,Anticor3,Anticor5,Anticor6}. 

Here, we show  that anti-correlated two-electron escape is not the only feature that distinguishes slingshot-NSDI  from the other NSDI pathways. The main NSDI pathways,  which are considered in the current study, are the direct pathway, RESI and the double delayed pathway. In the latter pathway both electrons ionize with a delay following recollision.  We show that slingshot-NSDI has very distinct fingerprints  in both energy and angular two-electron probability distributions. These features can be observed experimentally. Hence, this study paves the way for identifying slingshot-NSDI by kinematically complete experiments  that employ carrier envelope phase (CEP)-controlled near-single-cycle pulses   \cite{Anticor7,PhysRevA.83.013412,Camus,Kling1}.

\section*{Method}

We consider He driven by a near-single-cycle laser pulse at intensities 5$\times$10$^{14}$ W/cm$^2$ and 7$\times$10$^{14}$ W/cm$^2$ at 400 nm.
Both  intensities are  below the recollision threshold, which corresponds to an intensity of 8.6$\times$10$^{14}$ W/cm$^2$. The latter corresponds to  the maximum energy of the  electron returning to the core being equal to the energy   needed to transition  to the first excited state of the remaining ion. The maximum energy of the electron returning to the core is  3.17 $\mathcal{E}_{0}^2/(4\omega^2)$ \cite{Corkum}, which is equal to 23.7 eV at 5$\times$10$^{14}$ W/cm$^2$ and 33.1 eV at 7$\times$10$^{14}$ W/cm$^2$; $\mathcal{E}_{0}$ and $\omega$  are the strength and  frequency of the field.

 We use a laser field of the form 
\begin{equation}
  \vec{\mathcal{E}}(\mathrm{t}) = \mathcal{E}_0 \exp\left(-2\ln 2 \left(\frac{\mathrm{t}}{\tau}\right)^2 \right)\cos\left( \omega \mathrm{t} + \phi \right) \hat{z}, 
\label{eq1}
 \end{equation}
 where  $\phi$ is the  CEP, and $\tau=2$ fs is the full-width-half-maximum of the pulse duration in intensity. We employ atomic units, unless otherwise stated.

We use a three-dimensional (3D) semiclassical model  that is formulated in the framework of the dipole approximation \cite{Agapi_paper_2008}.  
 Previous successes of this model include verifying that electron backscattering from the nucleus accounts for the finger-like structure in NSDI of He driven by long laser pulses at higher intensities   \cite{Agapi_paper_2008}. This finger-like structure was predicted theoretically \cite{Taylorvshape} and  obtained experimentally \cite{vshape1,vshape2}. Moreover,   it was explained in a classical framework \cite{Agapi_paper_2008,Chinesevshape}. In addition, using this 3D model, we  investigated the direct versus the delayed pathway of NSDI for He driven by a long duration laser pulse at 400 nm \cite{Agapi2}. For intensities ranging from below- to above-the-recollision threshold,  we  achieved excellent agreement with fully ab-initio quantum mechanical calculations. In addition, using this model we obtained
  very good agreement with experimental results for several observables of NSDI  for Ar when driven by near-single-cycle laser pulses at 800 nm \cite{Agapi10}. These observables were obtained as a function of  CEP   for intensities ranging from  below- to above-the-recollision threshold. 

 In this model, one electron (recolliding) tunnel-ionizes  through the field-lowered Coulomb-barrier with a tunnel-ionization rate that is described by   the quantum mechanical Ammosov-Delone-Krainov (ADK) formula  \cite{A1,A2}.   We select the tunnel-ionization time, t$_{0}$,  using importance sampling \cite{importancesampling} in the time interval the field is present, that is, [-2$\tau$,2$\tau$]. The importance sampling distribution is given by the ADK ionization rate. The exit point of the recolliding electron  is along the laser-field direction  and is computed using parabolic coordinates \cite{parabolic1}.  The electron momentum   is taken to be equal to zero  along the laser field while  the transverse momentum  is given by a Gaussian distribution  \cite{A1,A2}.  The initially bound electron  is  described by a microcanonical distribution \cite{Abrimes}.  The weight of each classical trajectory i  that we propagate in time is given by 
 
\begin{eqnarray}
\mathrm{W_i=W_i^1\cdot W_i^2},
\end{eqnarray}
where 
\begin{eqnarray}
\mathrm{W_i^1}\propto \left(\frac{1}{|\vec{\mathcal{E}}(\mathrm{t_{0}})|}\right)^{\mathrm{2n^*-1}}\exp\left(-\frac{2\kappa^{3}}{3|\vec{\mathcal{E}}(\mathrm{t_{0}})|}\right)
\end{eqnarray}
is the ADK ionization rate\cite{A1,A2} at the time $\mathrm{t_{0}}$ of tunnel-ionization. The effective principal quantum number, $\mathrm{n^*}$, is given by $\mathrm{I_{p_{1}}=Z^2/2n^{*2}}$, while $\kappa=\sqrt{2\mathrm{I_{p_1}}}$ and $\mathrm{I_{p_1}}$ is the first ionization potential.
 The weight for electron 1 to have a transverse velocity equal to $\mathrm{v_\perp}$ at the time $\mathrm{t_{0}}$ is denoted by $\mathrm{W_i^2}$ and is given by
\begin{eqnarray}
\mathrm{W_i^2}\propto\frac{\mathrm{v_\perp}}{|\vec{\mathcal{E}}(\mathrm{t_{0}})|}\exp\left(-\frac{\mathrm{v^2_{\perp}\kappa}}{|\vec{\mathcal{E}}(\mathrm{t_{0}})|}\right).
\end{eqnarray}
  
 Once the initial conditions are specified at time $\mathrm{t_{0}}$, the position and momentum of each electron are propagated classically in time. We do so using  the three-body Hamiltonian of the two electrons with the nucleus kept fixed. 
 All Coulomb forces are accounted for: the interaction of each electron with the nucleus and the laser field and the electron-electron interaction are all included in the time propagation.  
  We also account for    the Coulomb singularity by using  regularized coordinates \cite{KS}. 
During the time propagation each electron is interacting with the nucleus with charge $\mathrm{Z=2}$. A trajectory is labeled as a doubly-ionized event if asymptotically,  i.e. $\mathrm{t\rightarrow \infty}$, the energies of both electrons are positive. 
The double ionization probability is given by
\begin{eqnarray}
\mathrm{P_{DI}=\frac{\sum_{i}^{N_{DI} } W_{i}}{ \sum_{i}^{N} W_{i}}}
\end{eqnarray}
where $\mathrm{N_{DI}}$ and N are the numbers of doubly-ionized and all events, respectively.

 We identify the main pathways of energy transfer  in each double ionization  event.  To do so we compute  the  time difference between the recollision time $\mathrm{t_{rec}}$ and the ionization time t$\mathrm{_{i}}$ of each electron, with  $\mathrm{i=1,2}$. We label the electron that ionizes first  as electron 1 and the one that ionizes last as electron 2.
  To identify these times, for each classical trajectory, we compute the pair potential energy and obtain its maximum which corresponds to the time of minimum approach of the two electrons. We define this time of minimum approach as the recollision time. Moreover, we define 
  the ionization time for each electron,  $\mathrm{t_{i}}$,  as the time when the compensated energy $\mathrm{(p_{x,i}^{2}+p_{y,i}^2+(p_{z,i}-\mathcal{A}(t))^2)/2-Z/r_{i}}$ becomes positive and remains positive thereafter \cite{Leopold}, with  $\mathrm{i=1,2}$ and $\mathrm{{\bf{p}}_{i}=p_{x,i}\hat{x}+p_{y,i}\hat{y}+p_{z,i}\hat{z}}$; $\mathcal{A}(\mathrm{t})$ is the vector potential and $Z=2$. 
 We compare the time difference between the recollision time  and the ionization time  of each electron with the time interval t$\mathrm{_{diff}}$ where the electron pair potential energy undergoes a sharp change due to recollision. For the laser field intensities considered in this work, we find t$\mathrm{_{diff}}$ to be  roughly equal to 1/8 laser cycle (T).  
We list in Table 1 the conditions satisfied by the   direct, delayed or double delayed double ionization events. We find that in  the delayed pathway,   the probability for electron 2 to be the recolliding electron  increases with  
decreasing intensity.

\begin{table}[h]
  $\Delta \mathrm{t_1}$=$\mathrm{t_{1}}$-$\mathrm{t_{rec}}$ \& $\Delta \mathrm{t_2}$=$\mathrm{t_{2}}$-$\mathrm{ t_{rec}}$ \\
  \vspace{0.2cm}
\begin{tabularx}{\columnwidth}{|X|X|X|}
 \hline
 \centering direct & \centering delayed & \centering double delayed \tabularnewline

 \hline
 \centering
$\begin{aligned}
\hspace{0.1cm} \Delta \mathrm{t_2} &< \mathrm{t_{diff}} \\
\hspace{0.1cm}\mathrm{t_1} &<\mathrm{t_2}
\end{aligned}
$

& 
\centering
$\begin{aligned}
\hspace{0.1cm}\Delta \mathrm{t_{1}} &< \mathrm{t_{diff}} \\ 
\hspace{0.1cm}\Delta \mathrm{t_{2}} &>\mathrm{t_{diff}}
\end{aligned}
 $
& 
\centering
$\begin{aligned}
\hspace{0.2cm}\Delta \mathrm{t_{1}} &> \mathrm{t_{diff}}\\ 
\hspace{0.2cm}\Delta \mathrm{t_{2}} &> \mathrm{t_{diff}}
\end{aligned}$ \tabularnewline

\hline

\end{tabularx}
 \caption{Conditions for energy transfer double ionization pathways. }

\label{Table2}
 
\end{table}

For the results presented in this work, we consider the intensities  5$\times$10$^{14}$ W/cm$^2$ and 7$\times$10$^{14}$ W/cm$^2$. For both intensities, 12 CEPs are considered ranging from $\phi=0^{\circ}$ to $\phi=330^{\circ}$ in steps of 30$^{\circ}$. For each $\mathrm{\phi}$, at  7$\times$10$^{14}$ W/cm$^2$ we obtain roughly 10$^{4}$ doubly-ionized events   as a result of running 500, 12-hour jobs, while at 5$\times$10$^{14}$ W/cm$^2$ we obtain 5$\times$10$^{3}$ doubly-ionized events as a result of running 4000 jobs, 12-hour jobs; one job corresponds to 1 CPU.  For the results  presented regarding total double ionization  the average has been taken over all CEPs for each intensity.

\section*{Results}
\subsection*{Slingshot-NSDI}
First, we identify the prevailing pathways of double ionization when He is driven by a 2 fs laser pulse at 400 nm and at intensities of 5$\times$10$^{14}$ W/cm$^2$ and 7$\times$10$^{14}$ W/cm$^2$. We find that the pathways prevailing NSDI ionization are the delayed and the direct ones at 7$\times$10$^{14}$ W/cm$^2$ and  the delayed and the double delayed ones at 5$\times$10$^{14}$ W/cm$^2$.  For the delayed pathway, 
 we note a significant change that takes place with decreasing intensity.  The mechanism that prevails switches from RESI to slingshot-NSDI. Since both RESI and slingshot-NSDI are mechanisms of the delayed pathway, in both mechanisms  electron 1 ionizes soon after  recollision and electron 2 transitions to an excited state of He$^{+}$. Slingshot-NSDI and RESI differ in the mechanism that underlies  the subsequent ionization  of electron 2. In RESI this mechanism involves electron 2 ionizing with the assistance of the laser field at later times at extrema of the field. In contrast, in slingshot-NSDI we find that electron 2 ionizes with the assistance of both the nucleus and the laser field. Next, we briefly outline some of the properties of slingshot-NSDI \cite{AgapiSlingshot}.

\begin{figure} [ht]
\centering
 \includegraphics[width=0.6\linewidth]{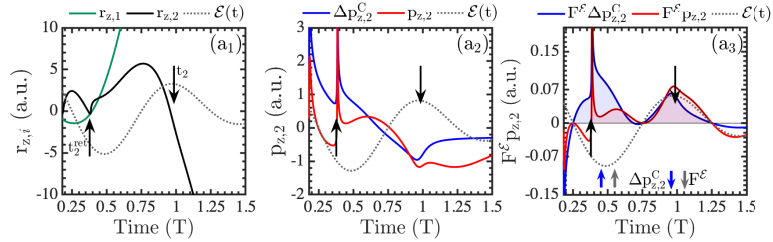}
\caption{Slingshot-NSDI. At  $5\times$10$^{14}$ W/cm$^2$, we plot as a function of time (a1) $\mathrm{r_{z,1}}$ and $\mathrm{r_{z,2}}$ (a2) $\mathrm{p_{z,2}}$ and $\mathrm{\Delta p_{z,2}^{C}}$ and (a3) F$\mathrm{^\mathcal{E} p_{z,2}}$ and F$\mathrm{^\mathcal{E} \Delta p_{z,2}}$. Coulomb slingshot is enclosed by the black arrows, with the up and down arrow depicting $\mathrm{p_{z,2}}$ 
being along  the +$\hat{z}$-axis and -$\hat{z}$-axis, respectively, at the start and the end of the Coulomb slingshot motion. The beginning of the time axis is t$\mathrm{_{rec}}$.}
\label{traj}
\end{figure}

In particular, \fig{traj}(a1)  shows that in slingshot-NSDI electron 2 undergoes a Coulomb slingshot motion around the nucleus,  see motion enclosed by the black arrows in \fig{traj}(a1). In \fig{traj}, we consider a double ionization event corresponding to CEP equal to zero but similar results hold for other CEPs. As a result of the slingshot motion, electron 2 escapes opposite to electron 1 and its momentum, p$_{z,2}$, undergoes a large change. The momentum of electron 2  is large and points along the direction of the force from the laser field at the start and at the end of the slingshot motion, see \fig{traj}(a2). 
 We  show that  this large change in the  momentum of electron 2  is due to the nucleus, by  expressing 
 the total momentum, p$_{z,2}$, as the sum of the momentum  changes  due to the interaction  with the nucleus and electron 1, $\mathrm{\Delta p_{z,2}^{C}}$,  and with  the laser field,  $\mathrm{\Delta p_{z,2}^{\mathcal{E}}}$ as follows:
\begin{subequations}
\begin{align}
\mathrm{p_{z,2}(t})&= \mathrm{p_{z,2}(t_{0})}+\mathrm{\Delta p_{z,2}^{C}(t_{0}\! \rightarrow \! t)} +\mathrm{\Delta p_{z,2}^{\mathcal{E}}(t_{0} \! \rightarrow \! t)} \\
\mathrm{\Delta p_{z,2}^{C}(t)} &= \mathrm{ \int_{t_{0}}^{t} \left(\frac{-Z \  r_{z,2}}{|r_2|^{3}} + \frac{r_{z,2}-r_{z,1}}{|{\bf{r}}_2 - {\bf{r}}_1|^{3}}\right)dt'}\\
\mathrm{\Delta p_{z,2}^{\mathcal{E}}(t)} &= \mathcal{A}(\mathrm{t}) - \mathcal{A}(\mathrm{t_{0}}). 
\label{eq:2}
\end{align}
\end{subequations}
 In the delayed pathway,  the  repulsive force between the two electrons is roughly zero shortly after recollision, thus,  contributing only  a  constant term  to $\mathrm{\Delta p_{z,2}^{C}}$.
We plot the momentum change due to the nucleus  of electron 2   in \fig{traj}(a2).
It is clear that, during the Coulomb slingshot motion,  the sharp change of the total momentum of electron 2      is mainly due  to the term $\mathrm{\Delta p_{z,2}^{C}}$.
 
 Next, we show that, during the slingshot motion, the laser field provides sufficient energy to  electron 2 to ionize due to the large change in the electron momentum.
 Shortly after recollision, at time  t$\mathrm{_{init}=t_{rec}+t_{diff}}$, the repulsive force between the two electrons is significantly smaller than during recollision. Hence, after this time, we can safely assume that the energy of electron 2 
changes due to the work done  mainly by the   field as follows:
\begin{equation}
\mathrm{H(t)=\frac{p_{2}(t_{init})^{2}}{2}  -\frac{Z}{r_{2}(t_{init})}+\int_{t_{init}}^{t}F^\mathcal{E}p_{z,2}dt'}, 
 \label{eq:3}
  \end{equation}
where  F$^\mathcal{E}(t)=-\mathcal{E}(t)$ is the force from the laser field and  $\mathrm{F^\mathcal{E}p_{z,2}}$ is the rate of change of the energy of  electron 2.
 During Coulomb slingshot, the close encounter of electron 2 with the nucleus at t$_{2}^{\mathrm{ret}}$   takes place past a zero of the laser  field. At this  time  both the momentum of electron 2   and the force from the laser field point along the +$\hat{z}$-axis. Roughly half a laser cycle later, in the time interval [0.75, 1.25]T, the slingshot motion is concluded with both the momentum of electron 2 and the force of the laser field pointing along the -$\hat{z}$-axis.  Thus, during  the Coulomb slingshot,  $\mathrm{F^\mathcal{E}p_{z,2}}$  is mostly positive in the first half cycle  [0.25, 0.75]T and the second one  [0.75, 1.25]T after recollision, see red-shaded area  in \fig{traj}(a3).   \fig{traj}(a3) clearly shows that 
 $\mathrm{F^\mathcal{E}p_{z,2}}$ is positive due to $\mathrm{F^\mathcal{E}\Delta p_{z,2}^{C}}$ being positive, see blue-shaded area in \fig{traj}(a3). Thus, the large change in the momentum of electron 2 due to Coulomb slingshot is the reason the rate of change of the energy provided to electron 2 by the  laser field is positive. This leads  to ionization of electron 2 around the second extremum of the field after recollision, i.e. in the time interval [0.75, 1.25]T. In contrast, in RESI electron 2 can ionize at any extremum of the laser field.

As we increase the laser intensity, it is more probable that the force exerted on electron 2 from the laser field is larger than the attractive Coulomb force exerted from the nucleus. As a result, in most delayed double ionization events, electron 2 does not return  at time $\mathrm{t_{2}^{ref}}$ close to the nucleus and therefore does not undergo Coulomb slingshot. That is, following the transition of electron 2 to an excited state,  the effect of the nucleus  on the motion of this electron decreases with increasing intensity. This results in  RESI becoming more important than slingshot-NSDI with increasing intensity.

\subsection*{Two-electron momentum probability distributions}
 
\begin{figure} [ht]
\centering
 \includegraphics[width=0.55\linewidth]{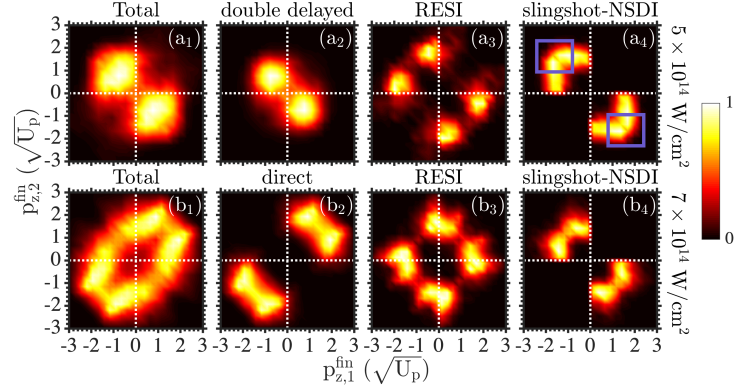}
\caption{Correlated electron momenta for NSDI pathways. We plot the correlated electron momenta at  intensities 5$\times$10$^{14}$ W/cm$^2$ (top row) and 7$\times$10$^{14}$ W/cm$^2$ (bottom row) for all double ionization events (a1) and (b1). At  5$\times$10$^{14}$ W/cm$^2$, we plot the correlated electron momenta for the double delayed (a2) and the delayed pathway, which includes RESI (a3) and slingshot-NSDI (a4). At 7$\times$10$^{14}$ W/cm$^2$, we plot the correlated electron momenta for the direct  (b2) and the delayed pathway,   which includes RESI (b3) and slingshot-NSDI (b4).}
\label{cor}
\end{figure}

  In \fig{cor}, we plot the correlated electron momenta. That is, we plot the double ionization probability as a function of each electron  momentum  along the direction of the laser field at intensities 5$\times$10$^{14}$ W/cm$^2$ (top row) and 7$\times$10$^{14}$ W/cm$^2$ (bottom row). As the intensity decreases, the two-electron escape changes  from correlated, with both electrons escaping in the same direction along the laser field  (\fig{cor}(b1)), to anti-correlated, see \fig{cor}(a1). Two are the main reasons for the transition from  anti-correlated two-electron escape at 5$\times$10$^{14}$ W/cm$^2$ to correlated one at 7$\times$10$^{14}$ W/cm$^2$. At both intensities we find that two pathways of NSDI prevail, with the delayed pathway being one of the two most important pathways in both cases.  However, the other prevailing pathway is the direct one at the higher intensity while it is the double delayed one at the lower intensity.  In the direct pathway, both electrons ionize soon after recollision around a zero of the laser field. As a result, the two electrons escape with large final electron momenta, roughly given by -$\mathcal{A}(\mathrm{t_{rec}})$, in the same direction along the laser field, see \fig{cor}(b2). In the double delayed pathway, the two electrons ionize with a delay after recollision resulting in smaller momenta along the laser field compared to direct events. In the double delayed events,  the electrons ionize mostly in opposite directions, see \fig{cor}(a2).  
  
  In addition, as the  intensity decreases from 7$\times$10$^{14}$ W/cm$^2$ to 5$\times$10$^{14}$ W/cm$^2$, the mechanism underlying the delayed pathway changes from RESI to slingshot-NSDI. At 7$\times$10$^{14}$ W/cm$^2$, once electron 2  transitions  to an excited state of the remaining ion, the electron subsequently ionizes mostly due to  the laser field around extrema of the field. As a result electron 2  escapes with a small final momentum. The electron that ionizes first escapes with large momentum, roughly equal to  -$\mathcal{A}(\mathrm{t_{rec}})$. Indeed, this is the pattern exhibited by the correlated electron momenta for RESI both at 5$\times$10$^{14}$ W/cm$^2$ and  at 7$\times$10$^{14}$ W/cm$^2$, see \fig{cor}(a3) and (b3). At 5$\times$10$^{14}$ W/cm$^2$, through Coulomb  slingshot, the nucleus plays a significant role in ionizing electron 2 after it transitions to an excited state. This effect results in both electrons escaping  with large momenta in opposite directions along the laser field,  see the two-electron probability distribution enclosed by the squares in \fig{cor}(a4). The role of the nucleus diminishes with increasing intensity  even for  slingshot-NSDI events. Indeed,  at the higher intensity, for slingshot-NSDI events, electron 2  does not have quite as large momentum as the first to ionize electron, see \fig{cor}(b4) and compare with \fig{cor}(a4).
  
\begin{figure} [ht]
\centering
 \includegraphics[width=0.55\linewidth]{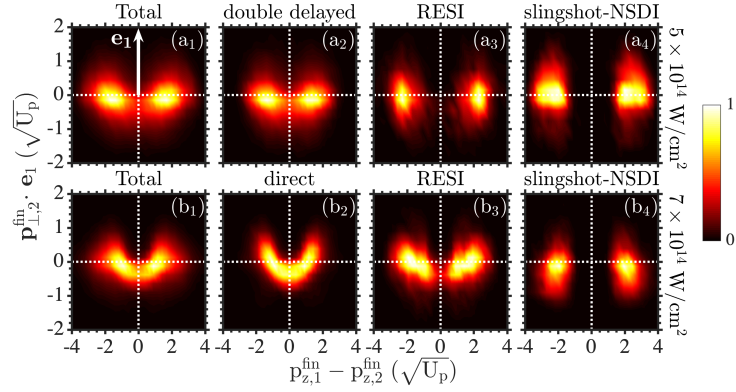}
\caption{Two-electron momenta distributions. The horizontal axis corresponds to the difference of the electron momenta along the laser field. The vertical axis is along the direction of the perpendicular to the laser field momentum of one of the two electrons, depicted by $\bf{e_{1}}$ in (a1). The vertical axis corresponds to the projection of the perpendicular momentum of the other electron on $\bf{e_{1}}$. We plot this two-electron probability distributions for all NSDI events as well as for different NSDI pathways at  5$\times$10$^{14}$ W/cm$^2$ (top row) and 7$\times$10$^{14}$ W/cm$^2$ (bottom row), as in \fig{cor}. }
\label{cor2}
\end{figure}

In \fig{cor2}, we plot the double ionization probability as a function of  the difference of the  electron momenta along the laser field and of the projection of the perpendicular   momentum of one electron  along the direction of the perpendicular momentum of the other electron, denoted by  $\bf{e}_{1}$ in \fig{cor2}(a1). At 7$\times$10$^{14}$ W/cm$^2$, \fig{cor2}(b1) shows that 
for NSDI events with similar electron momenta along the  laser field, i.e. $\mathrm{p_{z,1}^{fin}-p_{z,2}^{fin}\approx 0}$, electron-electron repulsion results in the two electrons escaping with opposite momenta in the direction perpendicular to the laser field. This pattern is due to the direct pathway of NSDI, as  \fig{cor2}(b2) clearly shows. In contrast, electron-electron repulsion does not
significantly contribute to RESI and slingshot-NSDI. This is  clearly seen in \fig{cor2}(b3) (RESI) and (b4) (slingshot-NSDI) at 7$\times$10$^{14}$ W/cm$^2$ and in \fig{cor2}(a3) (RESI) and (a4) (slingshot-NSDI) at 5$\times$10$^{14}$ W/cm$^2$. At the lower intensity, a comparison of \fig{cor2} (a2), (a3) and (a4) shows that electron-electron correlation plays a more important role for the double delayed events rather than for the RESI and the slingshot-NSDI events. This is consistent with both electrons ionizing with a delay in the double delayed events.The authors in ref. [23], use a similar two-electron distribution for all double ionization events for He driven by long laser pulses at small intensities at 400 nm. They find that electron-electron
 repulsion is smaller for  anti-correlated two-electron escape rather than for a correlated one. This is consistent with our finding that electron-electron correlation is larger for direct events compared to RESI and slingshot-NSDI events. 

\subsection*{Two-electron angular probability distributions}
\begin{figure} [ht]
\centering
 \includegraphics[width=0.55\linewidth]{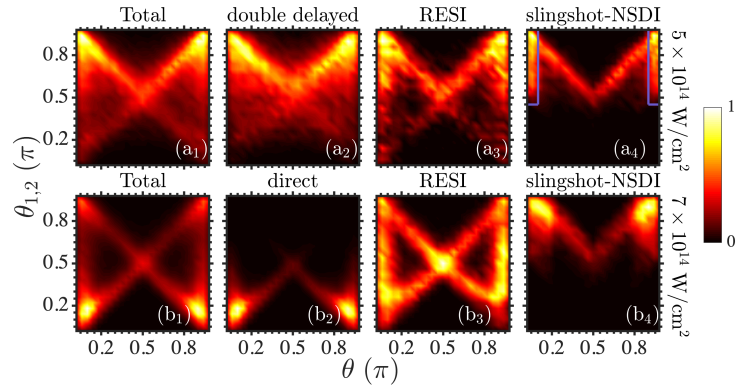}
\caption{Two-electron angular probability distributions. We plot the double ionization probability as a function of the inter-electronic angle  $\theta_{1,2}$ and as a function of the angle $\theta$  at  intensities 5$\times$10$^{14}$ W/cm$^2$ (top row) and 7$\times$10$^{14}$ W/cm$^2$ (bottom row) for all double ionization events (a1) and (b1) and for the dominant NSDI pathways  as in \fig{cor}.}
\label{angles}
\end{figure}

In \fig{angles}, we plot the double ionization probability as a function of the inter-electronic angle between the momenta of the two escaping electrons and the angle formed by the momentum of one of the two electrons  with respect to the axis of the  laser field. Comparing  \fig{angles}(a1) and (b1), we find  that  for most double ionization events the inter-electronic angle is close to 180$^{\circ}$
at  5$\times$10$^{14}$ W/cm$^2$ and close to 0${^\circ}$ at 7$\times$10$^{14}$ W/cm$^2$. This is  consistent with anti-correlated two-electron escape prevailing   at 5$\times$10$^{14}$ W/cm$^2$ and  correlated prevailing at 7$\times$10$^{14}$ W/cm$^2$, as we have already seen in \fig{cor}(a1) and (b1), respectively. 
At 5$\times$10$^{14}$ W/cm$^2$, \fig{angles}(a1) shows that for most double ionization events, in addition to the two electrons escaping opposite to each other, one of the two electrons escapes along the polarization direction, i.e. $\theta$ is 0$^{\circ}$ or 180$^{\circ}$. This latter pattern of two-electron escape  is more pronounced for slingshot-NSDI, see \fig{angles}(a4) and compare with \fig{angles}(a2) and (a3). We find that the electron escaping along the polarization axis is the one that ionizes first after recollision, giving rise to the probability distribution enclosed by the parallelograms in \fig{angles}(a4). The remaining wedge-like shape in \fig{angles}(a4) is accounted for by electron 2  forming an angle $\theta$ with the polarization axis while, for $\theta$ $\in[0^{\circ},90^{\circ}]$,  electron 1  forms an angle 180$^{\circ}$ giving rise 
 to $\theta_{1,2}=180^{\circ}-\theta$ and, for $\theta$ $\in[90^{\circ},180^{\circ}]$,  electron 1
forms an angle 0$^{\circ}$ giving rise to $\theta_{1,2}=\theta$. At the higher intensity, the prevailing correlated two-electron escape  is mostly due to the direct pathway of double ionization, see \fig{angles}(b2) and compare with \fig{angles}(b3) and (b4).

\subsection*{Two-electron energy probability distributions}

\begin{figure} [ht]
\centering
 \includegraphics[width=0.55\linewidth]{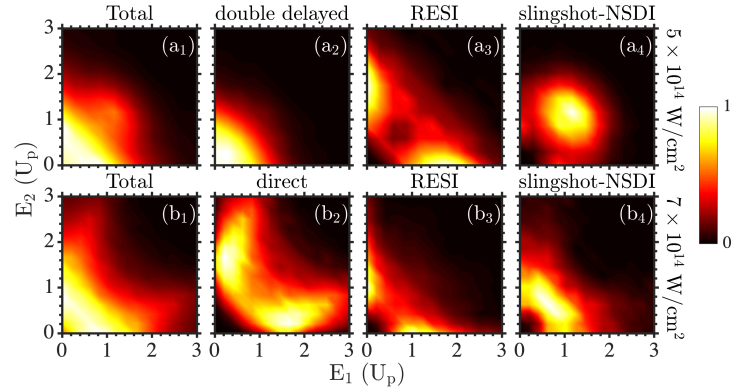}
\caption{Two-electron energy probability distributions. We plot the double ionization probability as a function of the energies of the two escaping electrons expressed in $\mathrm{U_{p}}$ at laser intensities 5$\times$10$^{14}$ W/cm$^2$ (top row) and 7$\times$10$^{14}$ W/cm$^2$ (bottom row) for all double ionization events (a1) and (b1) and for the dominant NSDI pathways as in \fig{cor}.}
\label{energy}
\end{figure}

Besides  anti-correlated two-electron escape, we find that another hallmark of slingshot-NSDI is both electrons escaping with large energy. This is clearly seen at  the smaller intensity of 5$\times$10$^{14}$ W/cm$^2$ in \fig{energy}(a4). In contrast, RESI gives rise to an unequal energy sharing between the two escaping 
electrons. This can be seen at 5$\times$10$^{14}$ W/cm$^2$ in \fig{energy}(a3) and at 7$\times$10$^{14}$ W/cm$^2$ in \fig{energy}(b3). Both in RESI and slingshot-NSDI electron 1 escapes with large energy. This is  mainly due to the large momentum of electron 1 along the direction of the laser field at the recollision time, which is roughly equal with -$\mathcal{A}(\mathrm{t_{rec}})$. The main difference between RESI and slingshot-NSDI is the influence of the nucleus on electron 2  following its transition to an excited state after recollision.  For RESI, the nucleus has a very small effect on electron 2. This  electron   ionizes with the help of the laser-field around field extrema, resulting in mostly small  final energy of electron 2. As a result,  unequal energy sharing 
prevails in RESI, see  \fig{energy}(a3) and (b3). In contrast, in slingshot-NSDI the nucleus plays a major role on electron 2. As we have previously discussed, this electron undergoes a Coulomb slingshot motion around the nucleus gaining a large amount of energy. As a  result, roughly equal energy sharing prevails in slingshot-NSDI,  see  \fig{energy}(a4). This pattern gives rise to 
a high concentration of NSDI probability around the diagonal at large energies in the two-electron energy probability distribution for all NSDI events, see \fig{energy}(a1).

The effect of the nucleus on electron 2, following its transition to an excited state, decreases with increasing intensity. This is manifested in RESI overtaking slingshot-NSDI as the prevailing mechanism of the delayed pathway. This diminishing effect of the nucleus on electron 2 in the delayed pathway  is also evident in slingshot-NSDI events.  Indeed, the two electrons share the energy more unequally at 7$\times$10$^{14}$ W/cm$^2$ in \fig{energy}(b4) compared to 5$\times$10$^{14}$ W/cm$^2$ in \fig{energy}(a4). The reduced contribution of slingshot-NSDI to the delayed pathway as well as the reduced effect of the nucleus on electron 2 in slingshot-NSDI at 7$\times$10$^{14}$ W/cm$^2$, accounts for the more unequal energy sharing of the two electrons for all NSDI events in \fig{energy}(b1) versus more equal energy sharing  at  5$\times$10$^{14}$ W/cm$^2$ in \fig{energy}(a1).  

\section*{Conclusions}
Using a 3D semiclassical model we investigate how two-electron probability distributions change for different NSDI pathways as well as for different  intensities below-the-recollision-threshold. We do so for He driven by near-single-cycle laser pulses at 400 nm. This study allows us to gain significant insight into slingshot-NSDI, which is a new mechanism 
that prevails the delayed pathway of NSDI at low intensities, as we have recently shown \cite{AgapiSlingshot}. We find that in slingshot-NSDI the electron that ionizes first does so along the polarization direction of the laser field. The electron that ionizes  last escapes opposite  to the other electron. This is due to electron 2 undergoing a Coulomb slingshot motion around the nucleus. Indeed, we have shown that, following recollision,  unlike direct events of NSDI where electron-electron correlation has a significant effect, electron repulsion has a very small effect on both RESI and slingshot-NSDI. Moreover, we find that in slingshot-NSDI the two electrons escape both with large energy giving rise to a distinct equal energy pattern in the two-electron energy probability distribution for all double ionization events. Thus, in addition to the anti-correlated two-electron escape, which we have previously identified as a hallmark 
of slingshot-NSDI \cite{AgapiSlingshot}, two-electron escape with roughly equal energy sharing and large energies is yet another trademark of slingshot-NSDI which can be measured by future experiments.

%

%

\section*{Acknowledgements}
A.E.  acknowledges the EPSRC grant no. N031326 and the use of the computational resources of Legion at UCL.

\section*{Author contributions statement}

G. P. Katsoulis performed the analysis of the computations involved and contributed to the ideas involved. A.E.  provided the codes used for the computations and the analysis and was responsible for the main ideas involved in the theoretical analysis. All authors reviewed the manuscript. 

\section*{Additional information}
\textbf{Competing financial interests:} The authors declare no competing financial interests.
%

\end{document}